# Topotactic growth of Zintl phase $Eu_5In_2As_6$ nanowires with antiferromagnetic behavior

Man Suk Song,[1] Lothar Houben,[2] Nadav Rothem,[3] Ambikesh Gupta,[1] Shai Rabkin,[3] Beena Kalisky,[3] Haim Beidenkopf,[1]* and Hadas Shtrikman[1]*

[1]Department of Condensed Matter Physics, Weizmann Institute of Science, Rehovot 7610001, Israel

[2]Department of Chemical Research Support, Weizmann Institute of Science, Rehovot, 7610001, Israel

[3]Department of Physics and Institute of Nanotechnology and Advanced Materials, Bar-Ilan University, Ramat Gan, 5290002, Israel

*E-mail: haim.beidenkopf@weizmann.ac.il, hadas.shtrikman@weizmann.ac.il

**ABSTRACT**

We demonstrate a topotactic transformation of zincblende InAs(Sb) nanowires into the Zintl phase $Eu_5In_2As_6$ through a vapor–solid mutual exchange process involving Eu and In in molecular beam epitaxy. This conversion preserves the polyhedral coordination lattice of the parent InAs(Sb) structure while inducing orthorhombic symmetry in the product phase, $Eu_5In_2As_6$, of which quasi-one-dimensional $[InAs_3]^{6-}$ chains with tetrahedral sites align along the <110> direction of zincblende structure. Local and global magnetic characterization identified two distinct antiferromagnetic phase transitions at approximately 7 K and 16 K in $Eu_5In_2As_6$ nanowires, potentially classified as altermagnetic phases. The versatility of the topotactic conversion of III-V semiconductor nanowires provides a platform for designing functional Zintl materials with tunable magnetic properties, making them promising candidates for spintronic applications.

III–V semiconductor nanowires (NWs) have emerged as fundamental building blocks in nanotechnology thanks to their exceptional electronic and optical properties. Indium arsenide (InAs) NWs, in particular, are notable for their high electron mobility and strong spin–orbit coupling, which make them ideal for advanced applications in electronics, photonics, and quantum computing.[1–4] One significant application involves hybridizing InAs NWs with superconductors to realize Majorana zero modes under an external magnetic field. The strong spin–orbit coupling and Zeeman splitting in InAs NWs enable the formation of these exotic quasiparticles, which are promising for topological quantum computing thanks to their non-Abelian statistics and potential for fault-tolerant computation.[5–8] Additionally, doping InAs NWs with small amounts of transition metals like manganese (Mn) and iron (Fe) induces ferromagnetic (FM) ordering, resulting in diluted magnetic semiconductors (DMS).[9,10] These DMS materials are promising candidates for spintronic devices, where the electron's spin degree of freedom is utilized for information processing and storage, potentially enhancing device functionality and efficiency.[11–13]

Recently, a topotactic growth mechanism using the rare-earth element europium (Eu) has been introduced to induce antiferromagnetic (AFM) ordering in InAs NWs. Specifically, solid-state mutual cation exchange has been used to convert wurtzite (WZ) InAs NWs into the Zintl phase

Eu$_3$In$_2$As$_4$, which exhibits AFM behavior.[14–16] Thus, this innovative approach provides the means for the incorporation of magnetic properties into the as-grown III-V materials. In this study, we extend the method of topotaxy to zincblende (ZB) InAs(Sb) NWs, demonstrating their conversion into the Zintl phase Eu$_5$In$_2$As$_6$, which orders antiferromagnetically below 17 K. The ability to transform the same semiconductor NWs into either FM or AFM materials through selective growth methods provides a versatile platform for designing spintronic devices with tailored magnetic properties. Such tunability could enable the development for future spintronic technologies.[17,18]

Vertical and reclining core NWs were grown by Au-assisted vapor–liquid–solid (VLS) molecular beam epitaxy (MBE) on (111)B and (001) InAs substrates, respectively. To obtain a ZB structure, we added a low flux of Sb after short growth of WZ InAs NWs in our previous work.[19] It was demonstrated, however, that the WZ InAs transforms into Zintl-phase Eu$_3$In$_2$As$_4$ shells via mutual-exchange topotaxy.[14] Without WZ InAs stumps, ZB InAs$_{1-x}$Sb$_x$ NWs (Sb 5–7 atomic %) were grown for 1.5 hours by opening both As and Sb shutters simultaneously under a high pressure As environment (Methods in SI). The ZB InAsSb cores are approximately 50 nm in diameter and 3–4 μm long. Subsequently, the In source was closed and cooled down, and Eu and As were evaporated for 2 hours to form the Eu-In-As shell. Scanning electron microscopy (SEM) images of the resulting NWs on both substrates are shown in Fig. 1a and 1b. Zoomed-in SEM images (Fig. 1c and 1d) show a polycrystalline-like structure with pointed shapes along the core NW axis. It is worth noting that adding Eu to the MBE chamber did not impede the growth of III–V NWs; however, the purity and the resulting mobilities might be degraded.

We examined the NW's stoichiometry and crystal structure by transmission electron microscopy (TEM). The crystallites grew both outwards as a shell and inwards penetrating into the NW core (Figs. S01–S03) as we observed in Eu$_3$In$_2$As$_4$ NWs[14]. The crystalline structure of the shells in TEM images has a different morphology compared to those seen along the Eu$_3$In$_2$As$_4$ NWs. Furthermore, energy-dispersive X-ray spectroscopy (EDS) analysis was performed to resolve the elemental composition. Mutual exchange of In from the core and Eu from the shell takes place across the boundary (yellow dashed lines in Fig. 1e) along the ZB NW, similar to the Eu$_3$In$_2$As$_4$ crystallites on WZ NWs. The atomically sharp phase boundary (light-blue colored dashed line in Fig. 1e) forms between the ZB core and the Eu-In-As shell (Fig. S04). From the analysis of the relative EDS intensities (Fig. S04–S05), we determined the stoichiometry of such shells to be Eu$_5$In$_2$As$_6$ despite Sb intervention in both core and shell. Eu$_5$In$_2$As$_6$ grains can also form via mutual exchange in reclined NWs on (001) substrates (approximately 55° relative to the substrate normal). The large angle at which Eu and As atoms impinge on the surfaces of these reclined NWs enhances the mutual exchange process compared to vertical NWs.[20] As a result, an axial junction of Eu$_5$In$_2$As$_6$ grains and core is formed (Fig. S06), making detailed investigation challenging. In the following discussion, we will focus solely on vertical NWs.

High-resolution scanning transmission electron microscopy (HRSTEM) demonstrated the perfectly ordered structure of the Eu$_5$In$_2$As$_6$ crystallites (Fig. S07). HRSTEM-EDS shown in Fig. 1h allowed us to elucidate the material structure as another Zintl phase with the orthorhombic space group *Pbam*.[21] Its unit cell is shown in Fig. 1g as well as overlaid in Fig. 1h. It includes two quasi-one-dimensional polyanionic chains of [InAs$_3$]$^{6-}$ aligned with the c-axis of the orthorhombic Eu$_5$In$_2$As$_6$ phase. The double chains are allocated to [In$_2$As$_6$]$^{10-}$ ribbons with As-As bonds, as described in Ref. 21 and 22. The Eu atoms are arranged among the [In$_2$As$_6$]$^{10-}$ ribbons in an intertwined pattern resembling the Greek letter lambda (λ), with each stroke containing five Eu$^{2+}$ cations[22] (Fig. S07d). This coordination is also characteristic of the isostructural Zintl phases of Ca$_5$Ga$_2$As$_6$ and Sr$_5$In$_2$As$_6$.[21,23] In a mechanistic picture, continuous chains of edge-sharing InAs tetrahedra in Eu$_3$In$_2$As$_4$ are formed from the corner-sharing tetrahedra in WZ InAs, and split into isolated anionic chains during the Eu uptake. The corner-sharing tetrahedral coordination in ZB InAs, on the other hand, become the tetrahedral chains in Eu$_5$In$_2$As$_6$ without a change of their sharing configuration throughout an influx of Eu.

In our previous work,[14] the Zintl phase Eu$_3$In$_2$As$_4$ was formed as a shell on underlying WZ InAs core via a mutual-exchange process. The formation of Eu$_5$In$_2$As$_6$ differs from that of Eu$_3$In$_2$As$_4$ in many respects. Eu$_5$In$_2$As$_6$ was formed with a shape of grainy crystallites on ZB InAsSb NWs rather than forming a shell. The crystallites grow and elongate in specific directions, having angles of approximately 35°, 145°, and rarely 90° from the $\langle 111 \rangle$ direction (growth axis) of NWs, as indicated by blue arrows in Figs. 2a–c. HRTEM was performed to investigate the interface between Eu$_5$In$_2$As$_6$ and ZB InAsSb NWs. As shown in Figs. 2d–i, three types of interfaces were found, designated as types A, B, and C, which denote different zone axes of Eu$_5$In$_2$As$_6$. Fast Fourier transform (FFT) analysis for each type is shown in Figure S01. In type A (Figs. 2d–e), the orientation between Eu$_5$In$_2$As$_6$ and InAsSb is $\langle 100 \rangle_{EuInAs} \parallel \langle 110 \rangle_{InAsSb}$ and the interfacial planes are $(00\bar{1})_{EuInAs} \parallel (00\bar{1})_{InAsSb}$. For type B, the orientation relationship is $\langle 010 \rangle_{EuInAs} \parallel \langle 11\bar{2} \rangle_{InAsSb}$; however, the interfacial planes for type B are ambiguous. The Eu$_5$In$_2$As$_6$ grains in both type A and type B have the same tiling angle and neighboring grains with another zone axis, $\langle 110 \rangle_{EuInAs}$, which corresponds to a viewpoint near the midpoint between the a- and b-axes (Figs. S02 and S03). For type C (Figs. 2h–i), the orientation is $\langle 001 \rangle_{EuInAs} \parallel \langle 110 \rangle_{InAsSb}$, and the interfacial planes are $\langle \bar{1}00 \rangle_{EuInAs} \parallel \langle 111 \rangle_{InAsSb}$. Eu$_5$In$_2$As$_6$ crystallites of type C were rarely found. However, wing-shaped and ~90°-tilted grains near the tip of the NW (the upper two arrows in Fig. 2a) are equivalent to type C. Figs. 2a and 2h have a rotating relationship of 90° around the NW growth direction, considering the morphological shape and crystallographic direction of the wing-shaped grain illustrated in Figure S08.

All types of $Eu_5In_2As_6$ crystallites exhibit anisotropic growth along the c-axis and have selective tilting angles with respect to the InAsSb NWs since the InAs tetrahedra in the NWs are the structural motifs that reappear in the $[InAs_3]^{6-}$ tetrahedral chains in $Eu_5In_2As_6$. In Fig. S09 (zone axis [110] of InAs, type A and C), the $[InAs_3]^{6-}$ chain consists of continuous corner-shared InAs tetrahedra aligned along the c-axis. We observed that the c-axis of $Eu_5In_2As_6$ aligns parallel to any of the equivalent ⟨110⟩ directions of InAsSb, as shown in Figs. 2e and 2i. Hence, the interface between ZB and $Eu_5In_2As_6$ in type B cannot be observed in the zone axis $[11\bar{2}]$ of the InAs tetrahedra (seemingly WZ) since both crystal structures overlap along the viewing direction. For the same reason, $Eu_5In_2As_6$ crystallites with $⟨110⟩_{EuInAs}$ or at some midpoints between the a- and b-axes were found. This tendency is maintained in an exceptional $Eu_5In_2As_6$ formation. In Figure 3, a few NWs grew along the ⟨110⟩ direction and displayed smoothly coated $Eu_5In_2As_6$ shells, as observed in both SEM and TEM analyses. These extraordinary NWs have potential for application of well-defined InAs-$Eu_5In_2As_6$ core-shell NWs and possibly the axial growth of $Eu_5In_2As_6$ NWs.

Topotaxial mutual-exchange growth involves the structural similarity between the parent and resulting crystals, where cations such as Eu and In mutually exchange in the As skeleton.[14] Therefore, it is necessary to examine the crystallographic resemblance between Zintl $Eu_5In_2As_6$ and ZB InAs(Sb), as well as to understand the processes of transformation and rearrangement. As our previous studies[19] have shown, Eu atoms are highly reactive[24] and can readily permeate into the ZB InAs(Sb) matrix. After Eu intrusion, the corner-sharing InAs tetrahedra in the ZB structure are maintained, as mentioned above; however, the tetrahedra are both corner- and edge-shared with $Eu^{2+}$ octahedra. Half of them are alternately inverted, and $Eu^{2+}$ octahedra are intercalated three-dimensionally within the ZB matrix. This distribution of Eu atoms resembles that of Eu in the (EuIn)As mosaic structure (Figure S14). As a result, the InAs tetrahedra in the ZB structure are transformed into the $[In_2As_6]^{10-}$ double chains along the c-axis in the orthorhombic $Eu_5In_2As_6$ structure. Common with the transition of WZ InAs into $Eu_3In_2As_4$ previously reported by us[14] is the topotaxial character of the transformation. The As framework of the parent phase is maintained in the sense that each As atom is a node in the polyhedral network of the product phase. Here, the cubic symmetry elements of the InAs(Sb) parent phase afford the resulting phase to have a specific symmetry resulting in a unique phase and composition. For each of the nodes in the As framework, there is a *one-to-one* relationship between the site in the parent phase and the site in the product phase (Fig. 4). The polyhedral symmetry spanned by the node sites is maintained in a slightly distorted form. The distance between As sites in the $Eu_5In_2As_6$ lattice is dilated on average by as little as 2% when compared with the As sublattice in the cubic InAs(Sb). Fig. 4a shows the cubic InAs in a <110> viewing direction with its two types of tetrahedral sites T+ and T–, of which one is occupied by In in the parent phase. Chains of T+ and T– sites in InAs along the <110> correlate with the $[InAs_3]^{6-}$ chains in $Eu_5In_2As_6$ along [001] seen in Fig. 4b. Initially unoccupied octahedral sites on InAs {111}-planes are filled by Eu to form the planes of alternating $EuAs_6$ octahedra in the $Eu_5In_2As_6$ lattice. A second type of Eu site is coordinated in prismatic geometry, originating from tetrahedral sites in the InAs. The alignment of the $[InAs_3]^{6-}$ chains with tetrahedral sites along the InAs ⟨110⟩ directions provides the intuitive understanding for the observed orientational relationships where the c-axis of $Eu_5In_2As_6$ coincides with any of the six equivalent ⟨110⟩ directions of the parent phase. It further supports such an alignment that the top-view pattern of $Eu_5In_2As_6$ crystallites (Fig. S15) manifests a three-fold symmetry.

Mechanistically the As(Sb) sublattice in InAs(Sb) forms a non-rigid framework for the transformation. The topotactic conversion occurs under indium-deficient conditions (i.e., without any external indium supply). We hypothesize that the transformation is initiated by a co-diffusion of Eu and In through vacant sites in the framework while maintaining charge neutrality. The gradient in chemical potential for the diffusion process is maintained at the interface to the gas-phase by the supply of Eu but also As for the consumption of In. Unlike in the mosaic structure reported before,[19] where indium was continuously supplied, here the indium shortage promotes In diffusion out of the InAs NW core, resulting in the formation of $Eu_5In_2As_6$ grains. In contrast, the external indium supply prevents the formation of $Eu_5In_2As_6$ and embeds EuAs sheets within polycrystalline InAs. Therefore, the critical indium concentration is a key factor in the topotaxial formation of $Eu_5In_2As_6$.

We used a combination of local scanning superconducting quantum interference device (SQUID) measurements and global SQUID magnetometry in a commercial magnetic property measurement system (MPMS). In local scanning SQUID of individual NWs we did not detect DC magnetization up to 1 m$\Phi_0$ at the location of the $Eu_5In_2As_6$ NWs alongside finite AC susceptibility, $\chi$, shown in Fig. 5a and b, respectively. The absence of a DC signal put a strict limit on the strength of magnetization associated with a ferromagnetic order down to our sensitivity, but the strength of the measured paramagnetic signal signified the presence of magnetic moments. We further examined the temperature dependence of the AC susceptibility signal across the NWs. The paramagnetic signal from two different NWs, shown in Fig. 5c (Fig. S16 in SI), exhibited a non-monotonic behavior. We found two local maxima, one at 7 K and a slightly fainter one at 15-17K. The cusp profile and lack of DC magnetization indicated the two transitions in the magnetic order are into antiferromagnetic phases. Interestingly, even though the intensity of the measured signal from each NW was slightly different, the magnetic phase transitions were consistent.

Therefore, to improve the signal-to-noise and better resolve the magnetic phase diagram of the $Eu_5In_2As_6$ NWs in temperature as well as in a magnetic field, we measured the global magnetization of millions of NWs harvested and deposited over a non-magnetic $Si/SiO_2$ substrate (Methods in SI) for MPMS. Individual magnetization curves taken along temperature sweeps at various applied magnetic fields are shown in Fig. 5d. At low magnetic fields, we resolved the two cusps in the magnetization at about 6 K and 16 K. This result is consistent with that obtained from bulk $Eu_5In_2As_6$[22]. Both shifted gradually to lower temperatures as the applied field increased. The resulting magnetic field-temperature phase diagram is shown in Fig. 5e by plotting in false color the derivative of the magnetization with temperature, dM/dT, to enhance the transition temperature. It thus hosts two distinct antiferromagnetic phases whose exact spin structures call for further investigation. Intriguingly, given the Eu substructure within Zintl $Eu_5In_2As_6$ those magnetic orders are more accurately classified as altermagnetic[25–27].

In this work, we realized Zintl phase $Eu_5In_2As_6$ NWs, which present clear AFM ordering, through a vapor–solid topotactic mutual exchange in MBE. This approach builds on our recently introduced method for growing Zintl $Eu_3In_2As_4$ NWs.[14] Our findings suggest that the ZB or WZ polytypes of indium pnictide core NWs could yield separate orthorhombic Zintl compounds, $X_5In_2Y_6$ or $X_3In_2Y_4$ (X = Sr, Ba, Yb; Y = As, Sb), each crystallizing in the *Pbam* or *Pnnm* space group, respectively. Furthermore, since ZB is the prevalent bulk structure in III–V crystals, our observation of topotactic conversion of ZB NWs may open the path to the application of this methodology to convert III-V substructure into Zintl films, potentially unlocking new possibilities in materials design and device integration.

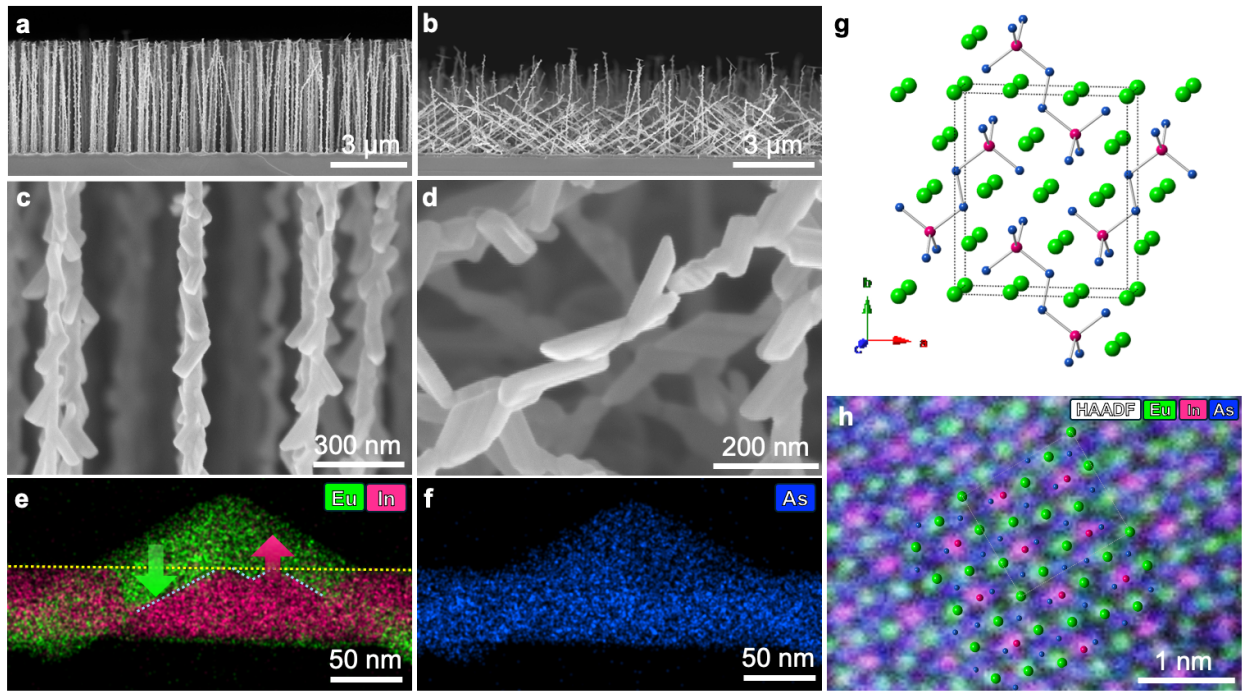

**Figure 1. Exchange growth of Zintl $Eu_5In_2As_6$ shell on ZB InAsSb core NWs.** (a, b) Cross-sectional SEM images of $Eu_5In_2As_6$ shell grown on ZB InAsSb core NWs, which were pre-grown on (111)B and (001) InAs substrates, respectively. (c, d) Magnified SEM images of $Eu_5In_2As_6$ NWs revealing grains with unique pointed shapes. (e) EDS maps of Eu (green) and In (magenta), showing their uniform distribution on either side of the core boundary. Green and magenta arrows indicate the mutual exchange of Eu and In atoms across the virtual boundary (yellow dashed line). (f) EDS map of As showing its uniform distribution across the core and shell. (g) Crystal structure of Zintl $Eu_5In_2As_6$. (h) EDS map of Eu (green), In (magenta), and As (blue) atoms overlaid on the high-resolution high-angle annular dark-field (HAADF) STEM image, showing the crystal structure of $Eu_5In_2As_6$ along the [001] zone axis.

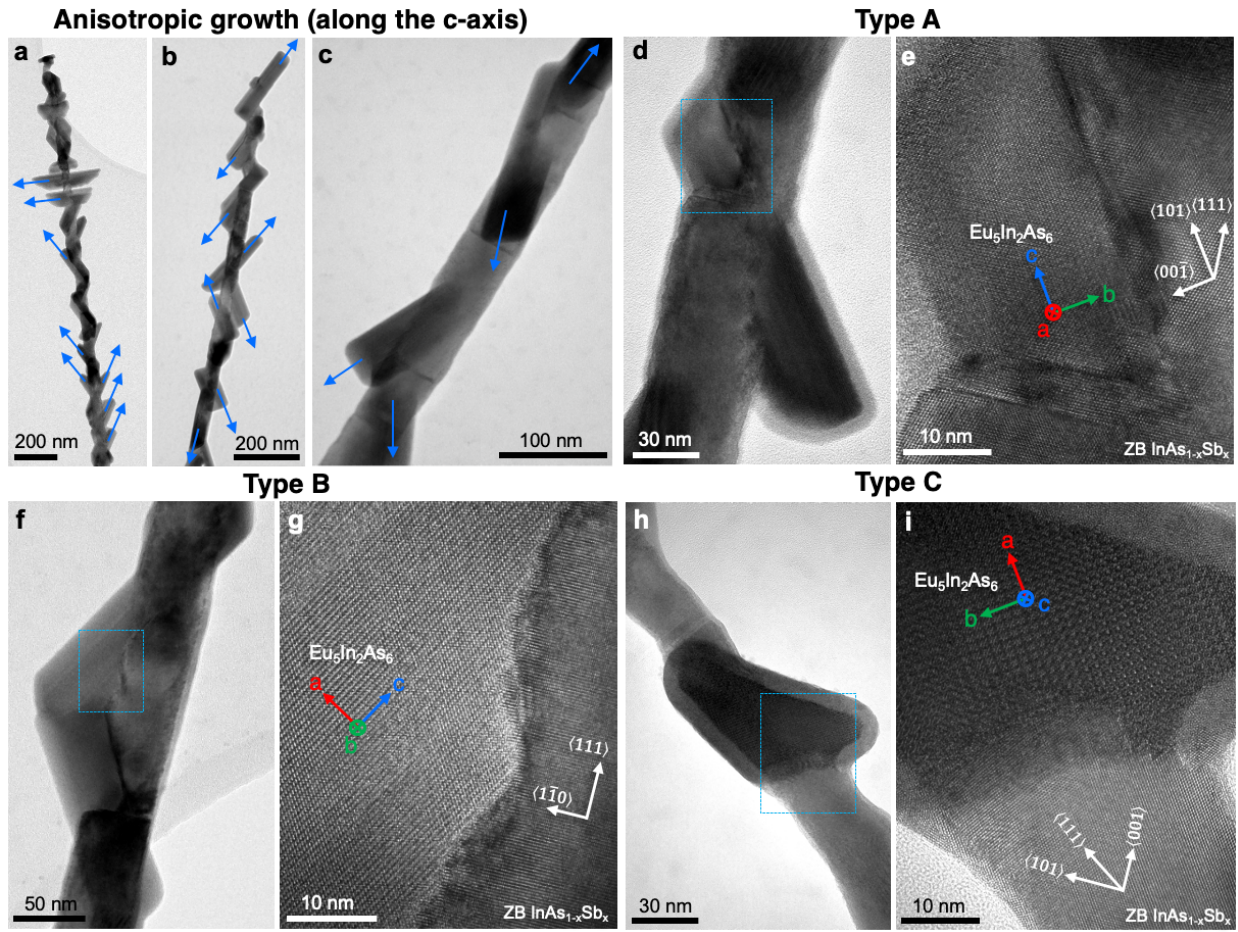

**Figure 2. Anisotropic growth of Zintl Eu$_5$In$_2$As$_6$ shell along the c-axis.** (a–c) TEM images of the as-grown Eu$_5$In$_2$As$_6$ on InAsSb NWs. The blue arrows indicate crystallographic c-axis direction of Eu$_5$In$_2$As$_6$, which forms angles of approximately 35°, 145°, rarely 90°, relative to the <111> direction of the core NWs. (d) Enlarged TEM image of Type A crystal. (e) HRTEM image of the area marked by a blue rectangle in (d), showing the interface in the <100>$_{EuInAs}$ and <110>$_{InAsSb}$ zone axes. (f) Enlarged TEM image of Type B crystal. (g) HRTEM image of the area marked by a blue rectangle in (f), showing the interface in the <010>$_{EuInAs}$ and <11$\bar{2}$>$_{InAsSb}$ zone axes. (h) Enlarged TEM image of Type C crystal. (i) HRTEM image of the area marked by a blue rectangle in (h), showing the interface in the <001>$_{EuInAs}$ and <110>$_{InAsSb}$ zone axes. This corresponds to the wing-shaped crystallites in (a) (the first and second blue arrows from the top). Respective fast Fourier transform (FFT) images used to identify the crystal structures and zone axes are shown in Figure S01.

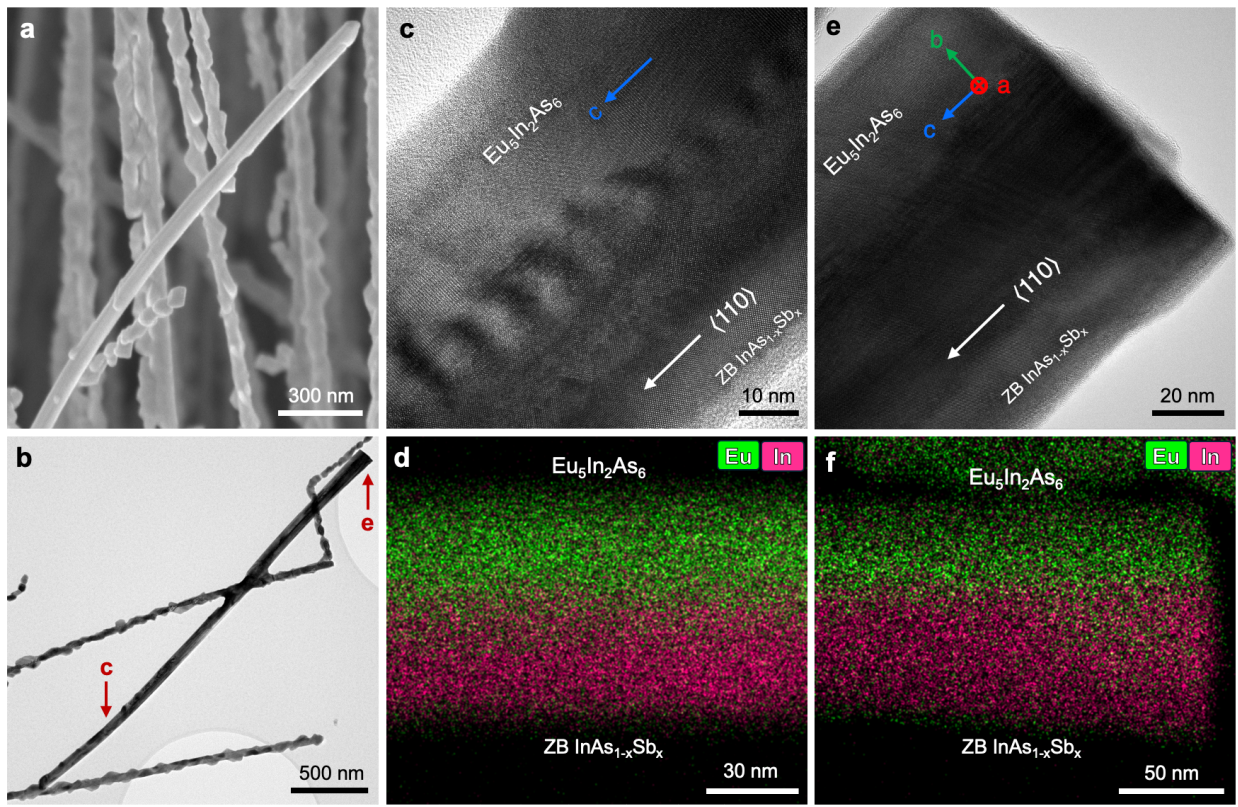

**Figure 3. Smoothly coated Eu$_5$In$_2$As$_6$ shell on <110>-oriented ZB InAsSb core NWs.** (a) SEM and (b) TEM images of as-grown Eu$_5$In$_2$As$_6$ on ZB InAsSb NWs grown exceptionally along the <110> direction. Enlarged TEM images near the tip (c) and the bottom (e), indicated by the red arrows in (b), show that the c-axis of Eu$_5$In$_2$As$_6$ is parallel to the <110> direction of the ZB core. (d, f) EDS elemental maps of Eu (green) and In (magenta) from (c) and (e), respectively. Only these two elements are shown to emphasize a different core-shell configuration. Detailed EDS and FFT analyses are provided in Figures S10-S13.

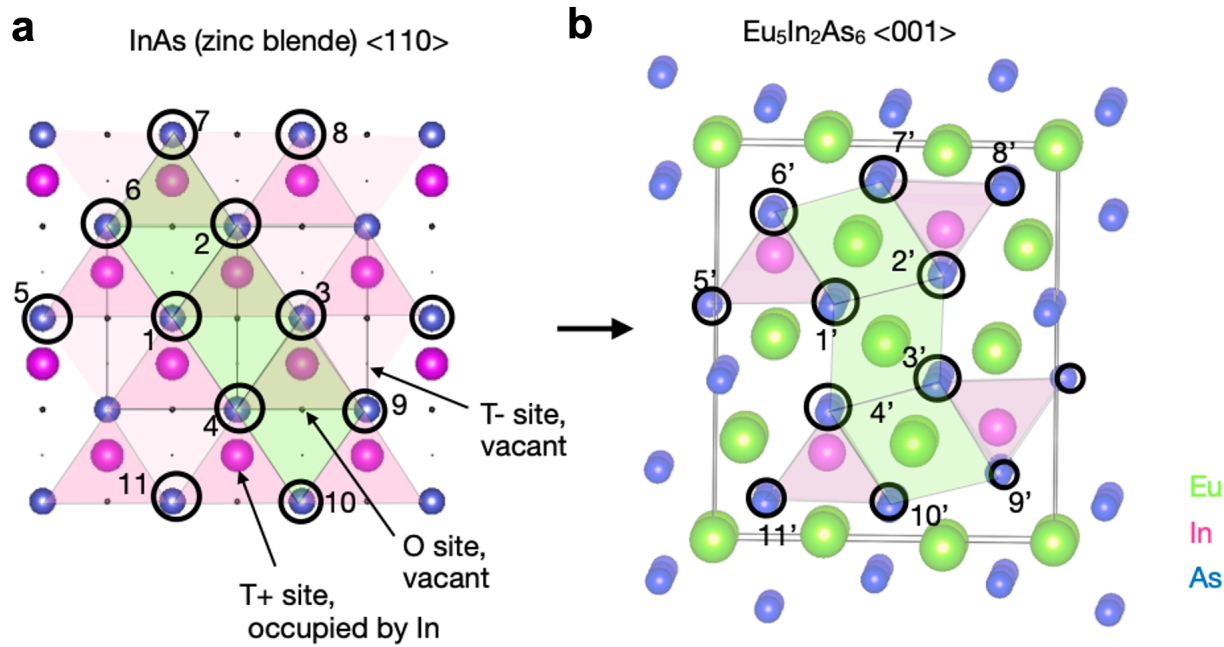

**Figure 4. Topotaxial transformation of zincblende InAs into orthorhombic Eu$_5$In$_2$As$_6$.** (a) Crystal structure of cubic InAs in <110> viewing direction. Tetrahedral sites T+ and T- are filled in pink, and octahedral sites are in green. (b) Crystal structure of Eu$_5$In$_2$As$_6$ in [001] viewing direction. Site nodes of the As framework map onto one another with a linear expansion of the lattice by about 2% in total. Layers of octahedrally coordinated Eu correspond to layers formed by filling vacant octahedral sites in the InAs on {111} planes (e.g. path 1-2-3-4 vs 1'-2'-3'-4'). Corner-sharing InAs tetrahedra aligned, and the related InAs tetrahedra forming [InAs$_3$]$^{6-}$ chains along [001] emerge from the tetrahedral T+ and T- sites along a <110> direction in the cubic InAs (paths 1-5-6 vs 1'-5'-6' and 2-7-8 vs 2'-7'-8').

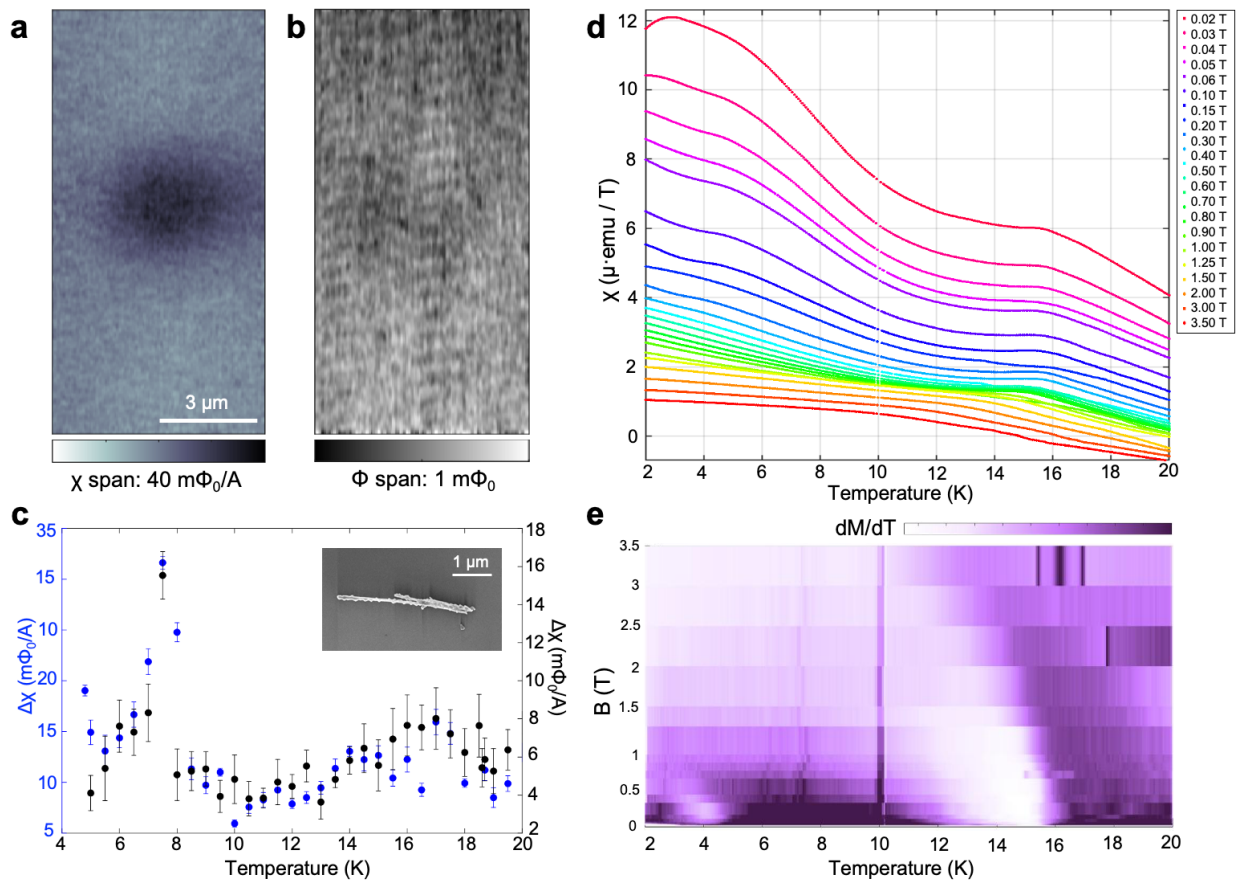

**Figure 5. Magnetic characterization of Zintl $Eu_5In_2As_6$ shell along the c-axis.** (a) Local AC susceptibility and (b) DC magnetism maps of $Eu_5In_2As_6$ NWs. (c) AC susceptibility of the NWs versus temperature, showing changes in trend at about 7 K and 15–17 K. The error bars are the standard deviation of pixel values in each relevant area in the vicinity of the two maximum local signals (Fig. S16). (d) Global susceptibility (c = B/H) measurements of the dispersed $Eu_5In_2As_6$ NWs with a SQUID in a MPMS showing the evolution of two Néel transition at about 6 K and 16 K. (e) Plot of dM/dT, where two Néel transition appear, presented in false color from (d).

ASSOCIATED CONTENT

**Supporting Information**. The Supporting Information is available free of charge.

It includes detailed experimental methods and additional structural, compositional, and magnetic characterization data of $Eu_5In_2As_6$ NWs. These data feature SEM, HR-TEM, STEM-HAADF,

EDS elemental mapping, schematic illustrations, and results from scanning SQUID measurements (PDF).

## Author Contributions

M.S.S., H.S. and H.B. conceptualized this research. H.S. and M.S.S. designed and carried out the MBE growths. M.S.S. and L.H. conducted the electron microscopy imaging with EDS analysis and modeling of the interfacial crystal structure. N.R., S.R., A.G., H.B. and B.K. performed magnetic measurements and analyzed the data. M.S.S., L.H. and H.B. wrote the manuscript. M.S.S., L.H. and H.B. prepared the figures in the main text and Supporting Information. All authors contributed to the discussion and manuscript preparation.

## Notes

M.S.S., L.H., H.S. and H.B. are inventors on US provisional patent no. 63/671,293 filed on 29 January 2024 under Yeda Research and Development Co. Ltd. The other authors declare no competing interests.

ACKNOWLEDGMENT

We deeply thank Michael Fourmansky and Israel Boterashvili for their professional assistance in the MBE growth of NWs and Markus Huecker in the MPMS measurements. We thank Katya Rechav for FIB lamella preparation and the consistent support of the Electron Microscopy Unit. L.H. acknowledges the support of the Irving and Cherna Moskowitz Center for Nano and Bio-Imaging at the Weizmann Institute of Science. H.S. is an incumbent of the Henry and Gertrude F. Rothschild Research Fellow Chair. N.R., S.R., and B.K. acknowledge the support of the European Research Council grant no. ERC-866236. H.B. and H.S. acknowledge support from the European Research Council (ERC-PoC TopoTapered–101067680), the Israeli Science

Foundation (grant 1152/23), and the Institute for Environmental Sustainability (IES) (grant no.

147977).

REFERENCES

(1) Lieber, C. M.; Wang, Z. L. Functional Nanowires. *MRS Bulletin* **2007**, *32* (2), 99–108. https://doi.org/10.1557/mrs2007.41.
(2) Yan, R.; Gargas, D.; Yang, P. Nanowire Photonics. *Nature Photon* **2009**, *3* (10), 569–576. https://doi.org/10.1038/nphoton.2009.184.
(3) Dayeh, S. A. Electron Transport in Indium Arsenide Nanowires. *Semicond. Sci. Technol.* **2010**, *25* (2), 024004. https://doi.org/10.1088/0268-1242/25/2/024004.
(4) Schroer, M. D.; Petta, J. R. Correlating the Nanostructure and Electronic Properties of InAs Nanowires. *Nano Lett.* **2010**, *10* (5), 1618–1622. https://doi.org/10.1021/nl904053j.
(5) Alicea, J.; Oreg, Y.; Refael, G.; von Oppen, F.; Fisher, M. P. A. Non-Abelian Statistics and Topological Quantum Information Processing in 1D Wire Networks. *Nature Physics* **2011**, *7* (5), 412–417. https://doi.org/10.1038/nphys1915.
(6) Mourik, V.; Zuo, K.; Frolov, S. M.; Plissard, S. R.; Bakkers, E. P. a. M.; Kouwenhoven, L. P. Signatures of Majorana Fermions in Hybrid Superconductor-Semiconductor Nanowire Devices. *Science* **2012**, *336* (6084), 1003–1007. https://doi.org/10.1126/science.1222360.
(7) Das, A.; Ronen, Y.; Most, Y.; Oreg, Y.; Heiblum, M.; Shtrikman, H. Zero-Bias Peaks and Splitting in an Al–InAs Nanowire Topological Superconductor as a Signature of Majorana Fermions. *Nature Physics* **2012**, *8* (12), 887–895. https://doi.org/10.1038/nphys2479.
(8) Sarma, S. D.; Freedman, M.; Nayak, C. Majorana Zero Modes and Topological Quantum Computation. *npj Quantum Information* **2015**, *1* (1), 1–13. https://doi.org/10.1038/npjqi.2015.1.
(9) Galicka, M.; Buczko, R.; Kacman, P. Structure-Dependent Ferromagnetism in Mn-Doped III–V Nanowires. *Nano Lett.* **2011**, *11* (8), 3319–3323. https://doi.org/10.1021/nl201687q.
(10) Nakamura, T.; Anh, L. D.; Hashimoto, Y.; Ohya, S.; Tanaka, M.; Katsumoto, S. Evidence for Spin-Triplet Electron Pairing in the Proximity-Induced Superconducting State of an Fe-Doped InAs Semiconductor. *Phys. Rev. Lett.* **2019**, *122* (10), 107001. https://doi.org/10.1103/PhysRevLett.122.107001.
(11) Ohno, H. Making Nonmagnetic Semiconductors Ferromagnetic. *Science* **1998**, *281* (5379), 951–956. https://doi.org/10.1126/science.281.5379.951.
(12) Awschalom, D. D.; Kawakami, R. K. Teaching Magnets New Tricks. *Nature* **2000**, *408* (6815), 923–924. https://doi.org/10.1038/35050194.
(13) Wolf, S. A.; Awschalom, D. D.; Buhrman, R. A.; Daughton, J. M.; von Molnár, S.; Roukes, M. L.; Chtchelkanova, A. Y.; Treger, D. M. Spintronics: A Spin-Based Electronics Vision for the Future. *Science* **2001**, *294* (5546), 1488–1495. https://doi.org/10.1126/science.1065389.

(14) Song, M. S.; Houben, L.; Zhao, Y.; Bae, H.; Rothem, N.; Gupta, A.; Yan, B.; Kalisky, B.; Zaluska-Kotur, M.; Kacman, P.; Shtrikman, H.; Beidenkopf, H. Topotaxial Mutual-Exchange Growth of Magnetic Zintl Eu3In2As4 Nanowires with Axion Insulator Classification. *Nat. Nanotechnol.* **2024**, *19* (12), 1796–1803. https://doi.org/10.1038/s41565-024-01762-7.

(15) Taha, T. A.; Mehmood, S.; Ali, Z.; Khan, S.; Aman, S.; Farid, H. M. T.; Trukhanov, S. V.; Zubar, T. I.; Tishkevich, D. I.; Trukhanov, A. V. Structure, Magnetic, Opto-Electronic and Thermoelectric Properties of A3In2As4 and A5In2As6 (A = Sr and Eu) Zintl Phase Compounds. *Journal of Alloys and Compounds* **2023**, *938*, 168614. https://doi.org/10.1016/j.jallcom.2022.168614.

(16) Jia, K.; Yao, J.; He, X.; Li, Y.; Deng, J.; Yang, M.; Wang, J.; Zhu, Z.; Wang, C.; Yan, D.; Feng, H. L.; Shen, J.; Luo, Y.; Wang, Z.; Shi, Y. Discovery of a Magnetic Topological Semimetal Eu3In2As4 with a Single Pair of Weyl Points. arXiv: 2403.07637v1 [cond-mat.mes-hall] (Submitted March 12, 2024). https://doi.org/10.48550/arXiv.2403.07637 (accessed 2024-12-31).

(17) Jungwirth, T.; Marti, X.; Wadley, P.; Wunderlich, J. Antiferromagnetic Spintronics. *Nature Nanotech* **2016**, *11* (3), 231–241. https://doi.org/10.1038/nnano.2016.18.

(18) Baltz, V.; Manchon, A.; Tsoi, M.; Moriyama, T.; Ono, T.; Tserkovnyak, Y. Antiferromagnetic Spintronics. *Rev. Mod. Phys.* **2018**, *90* (1), 015005. https://doi.org/10.1103/RevModPhys.90.015005.

(19) Shtrikman, H.; Song, M. S.; Załuska-Kotur, M. A.; Buczko, R.; Wang, X.; Kalisky, B.; Kacman, P.; Houben, L.; Beidenkopf, H. Intrinsic Magnetic (EuIn)As Nanowire Shells with a Unique Crystal Structure. *Nano Lett.* **2022**, *22* (22), 8925–8931. https://doi.org/10.1021/acs.nanolett.2c03012.

(20) Kang, J.-H.; Grivnin, A.; Bor, E.; Reiner, J.; Avraham, N.; Ronen, Y.; Cohen, Y.; Kacman, P.; Shtrikman, H.; Beidenkopf, H. Robust Epitaxial Al Coating of Reclined InAs Nanowires. *Nano Lett.* **2017**, *17* (12), 7520–7527. https://doi.org/10.1021/acs.nanolett.7b03444.

(21) Childs, A. B.; Baranets, S.; Bobev, S. Five New Ternary Indium-Arsenides Discovered. Synthesis and Structural Characterization of the Zintl Phases Sr3In2As4, Ba3In2As4, Eu3In2As4, Sr5In2As6 and Eu5In2As6. *Journal of Solid State Chemistry* **2019**, *278*, 120889. https://doi.org/10.1016/j.jssc.2019.07.050.

(22) Radzieowski, M.; Stegemann, F.; Klenner, S.; Zhang, Y.; Fokwa, B. P. T.; Janka, O. On the Divalent Character of the Eu Atoms in the Ternary Zintl Phases Eu5In2Pn6 and Eu3MAs3 (Pn = As–Bi; M = Al, Ga). *Mater. Chem. Front.* **2020**, *4* (4), 1231–1248. https://doi.org/10.1039/C9QM00703B.

(23) Verdier, P.; L'Haridon, P.; Maunaye, M.; Laurent, Y. Etude Structurale de Ca5Ga2As6. *Acta Cryst B* **1976**, *32* (3), 726–728. https://doi.org/10.1107/S0567740876003889.

(24) McWhan, D. B.; Souers, P. C.; Jura, G. Magnetic and Structural Properties of Europium Metal and Europium Monoxide at High Pressure. *Phys. Rev.* **1966**, *143* (2), 385–389. https://doi.org/10.1103/PhysRev.143.385.

(25) Šmejkal, L.; MacDonald, A. H.; Sinova, J.; Nakatsuji, S.; Jungwirth, T. Anomalous Hall Antiferromagnets. *Nat Rev Mater* **2022**, *7* (6), 482–496. https://doi.org/10.1038/s41578-022-00430-3.

(26) Šmejkal, L.; Sinova, J.; Jungwirth, T. Emerging Research Landscape of Altermagnetism. *Phys. Rev. X* **2022**, *12* (4), 040501. https://doi.org/10.1103/PhysRevX.12.040501.

(27) Krempaský, J.; Šmejkal, L.; D'Souza, S. W.; Hajlaoui, M.; Springholz, G.; Uhlířová, K.; Alarab, F.; Constantinou, P. C.; Strocov, V.; Usanov, D.; Pudelko, W. R.; González-Hernández, R.; Birk Hellenes, A.; Jansa, Z.; Reichlová, H.; Šobáň, Z.; Gonzalez Betancourt, R. D.; Wadley, P.; Sinova, J.; Kriegner, D.; Minár, J.; Dil, J. H.; Jungwirth, T. Altermagnetic Lifting of Kramers Spin Degeneracy. *Nature* **2024**, *626* (7999), 517–522. https://doi.org/10.1038/s41586-023-06907-7.

SYNOPSIS

Topotaxial transformation of zincblende InAs(Sb) structure into orthorhombic $Eu_5In_2As_6$.

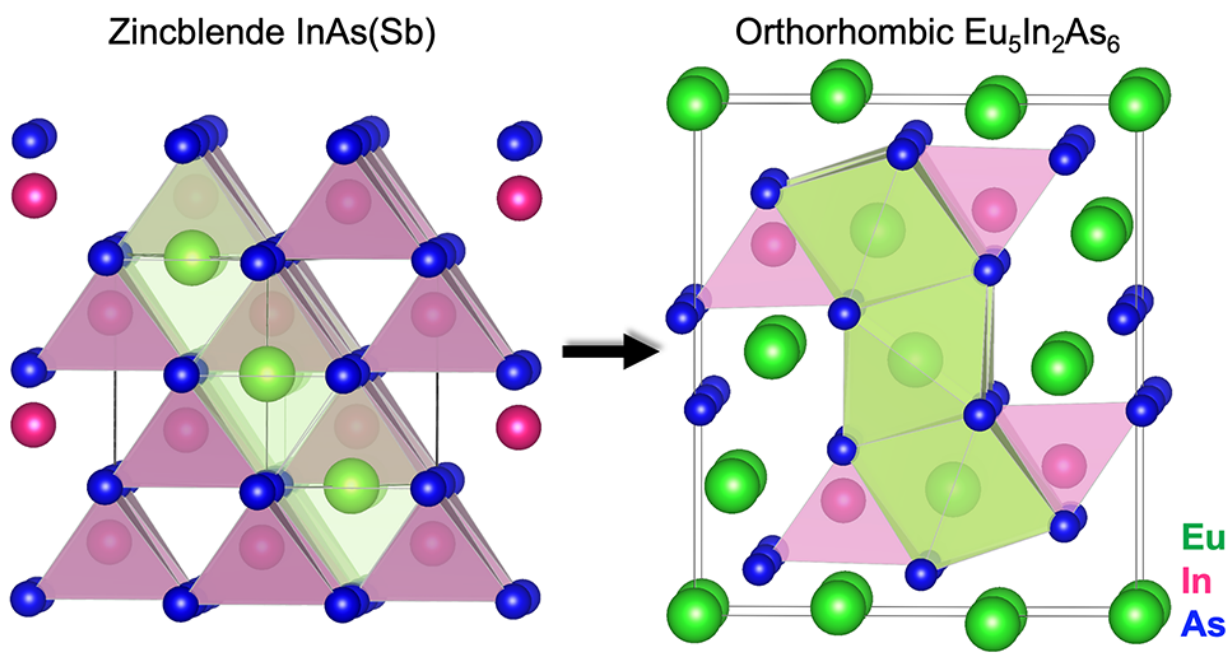

SUPPLEMENTARY INFORMATION

# Topotactic growth of Zintl phase Eu$_5$In$_2$As$_6$ nanowires with antiferromagnetic behavior

Man Suk Song,[1] Lothar Houben,[2] Nadav Rothem,[3] Ambikesh Gupta,[1] Shai Rabkin,[3] Beena Kalisky,[3] Haim Beidenkopf,[1]* and Hadas Shtrikman[1]*

[1]Department of Condensed Matter Physics, Weizmann Institute of Science, Rehovot 7610001, Israel

[2]Department of Chemical Research Support, Weizmann Institute of Science, Rehovot, 7610001, Israel

[3]Department of Physics and Institute of Nanotechnology and Advanced Materials, Bar-Ilan University, Ramat Gan, 5290002, Israel

## Methods

**Eu$_5$In$_2$As$_6$ NW growth**

Zintl Eu$_5$In$_2$As$_6$ grains were topotaxially grown on the surface of ZB InAs NWs. As a preliminary step, vertical and reclining InAs$_{1-x}$Sb$_x$ NWs were grown by MBE (RIBER-32 CBE) on (111)B and (001) InAs substrates, respectively, using the Au-assisted VLS techniques, as described in previous work[1–4]. InAs$_{1-x}$Sb$_x$ NWs (Sb 5–7 atomic %) were grown for 1.5 hours by opening both As and Sb shutters simultaneously under a high pressure As environment. In and Sb temperatures (and fluxes) were 720°C (1.1×10$^{-7}$ torr) and 425 °C (1.7×10$^{-7}$ torr), respectively. The growth temperature was maintained at the typical for the growth of InAs NWs temperature (410 °C). To initiate the topotaxial cation exchange, the Eu shutter was opened as the In cell was cold and its shutter closed at the As flux was maintained. A 15-min pause for adjusting the cell temperatures followed the InAs$_{1-x}$Sb$_x$ NW growth. Right after opening of the Eu shutter, the substrate temperature was ramped to a temperature 470 °C at a rate of 10 °C min$^{–1}$. The topotaxial exchange growth was maintained for 2 h. The temperatures (and fluxes) of Eu and As were 450 °C (3.8×10$^{–8}$ torr) and 222 °C ($1.8 \times 10^{-6}$ torr), respectively. The substrate manipulator in the MBE system was maintained at the RIBER's so-called "standard epitaxy position" during both steps, without any adjustments.

**Microscopy**

The Zintl $Eu_5In_2As_6$ NWs were characterized by field-emission SEM (Zeiss Supra-55, 3 kV, working distance of ~4 mm) and TEM (Thermo Fisher Scientific Talos F200X, 200 kV). The EDS composition data and mapping images were obtained by TEM with an attached detector that is identical to the one used in scanning transmission electron microscopy (STEM).

HRSTEM images and analytical EDS maps were acquired in a double-aberration-corrected Themis-Z microscope (Thermo Fisher Scientific Electron Microscopy Solutions) at an accelerating voltage of 200 kV. The STEM images were recorded with a Fischione Model 3000 detector and a Thermo Fisher Scientific bright field detector. The EDS hyperspectral data were obtained with a Super-X SDD detector and quantified with Velox software (Thermo Fisher Microscopy Solutions), version 3.13, through background subtraction and spectrum deconvolution. The STEM images were obtained with an electron probe with a convergence angle of 21 mrad and a primary beam current of less than 50 pA; the EDS maps were recorded at a beam current of 200 pA.

**SQUID**

We used scanning SQUID microscopy to search for magnetic signals from the NWs. A SQUID converts magnetic flux to voltage, allowing a sensitive detection of magnetic fields[5]. The planar SQUID, in the scanning configuration, allows the mapping of the static magnetic landscape and the local susceptibility[6]. The sensitive area of the SQUID used in this work—the pickup loop—has a diameter of 1.5 μm. Local susceptibility was measured using an on-chip coil to apply a magnetic field (the field coil, operated in this work at ~kilohertz and ~Gauss), whereas the pickup loop records the local response to the applied magnetic field. The positive signal in our data indicates a paramagnetic response. These measurements of susceptibility were plotted in units of $\Phi_0$ normalized by the sensitive area and current in the field coil.

We then performed these measurements as a function of temperature between 5 and 20 K to characterize the magnetic behaviour of the NWs. We later plotted the data as the inverse of the change in susceptibility.

In addition, the commercial SQUID magnetometer (MPMS3, Quantum Design) was also used to measure the massive $Eu_5In_2As_6$ NWs. To eliminate any possible magnetic contribution from the InAs substrate, we harvest the NWs by sonicating them for a few seconds in a solvent. We then disperse them over a $Si/SiO_2$ substrate ($5 \times 6$ mm$^2$) by depositing the suspension and drying it drop by drop to exhaust all the NWs available[7]. This sample was inserted into a straw and mounted in MPMS3. The global magnetic measurement was carried out in the temperature range between 2 and 20 K by varying the magnetic field from 0.02 to 3.50 T.

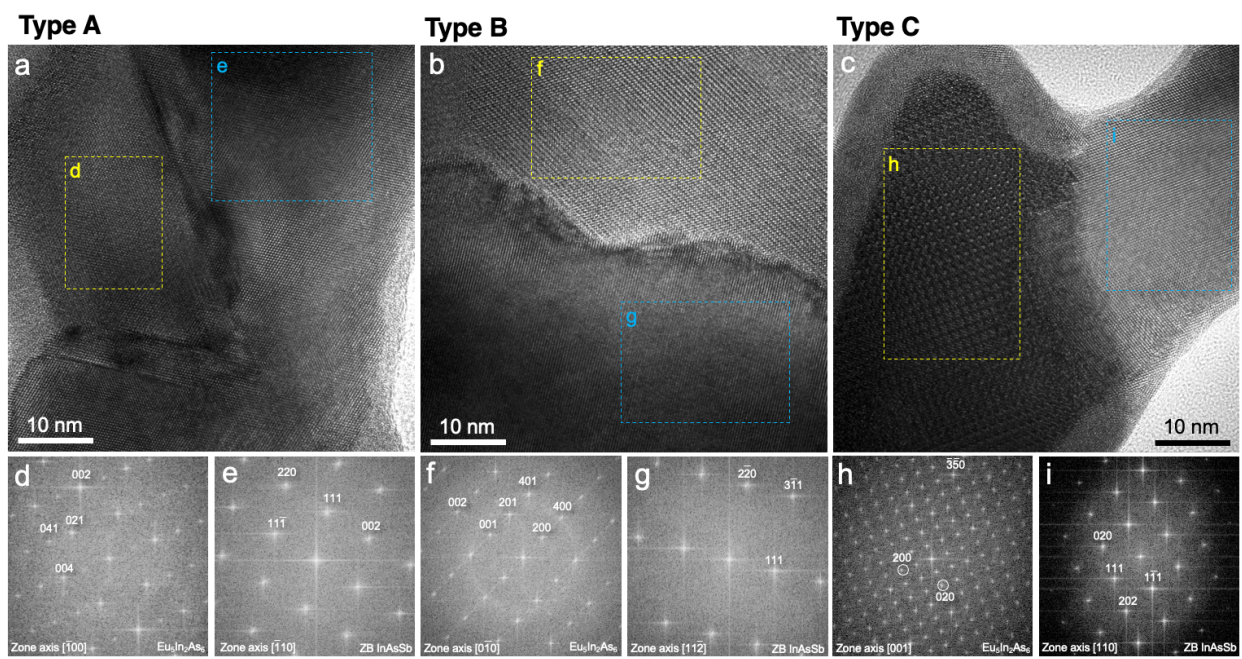

**Figure S01 HR-TEM and FFT analyses of Zintl Eu$_5$In$_2$As$_6$ grains for Type A, B, and C.** (a–c) Magnified HR-TEM images in Figures. 2e, 2g, and 2i, respectively. (d, f, g) FFT patterns of the yellow dashed rectangular areas in (a–c), demonstrating Eu$_5$In$_2$As$_6$ along the [-100], [0-10], and [001] zone axes, respectively. (e, g, h) FFT patterns of the blue dashed rectangular areas in (a–c), demonstrating ZB InAs(Sb) in the [-110], [11-2], and [110] zone axes, respectively.

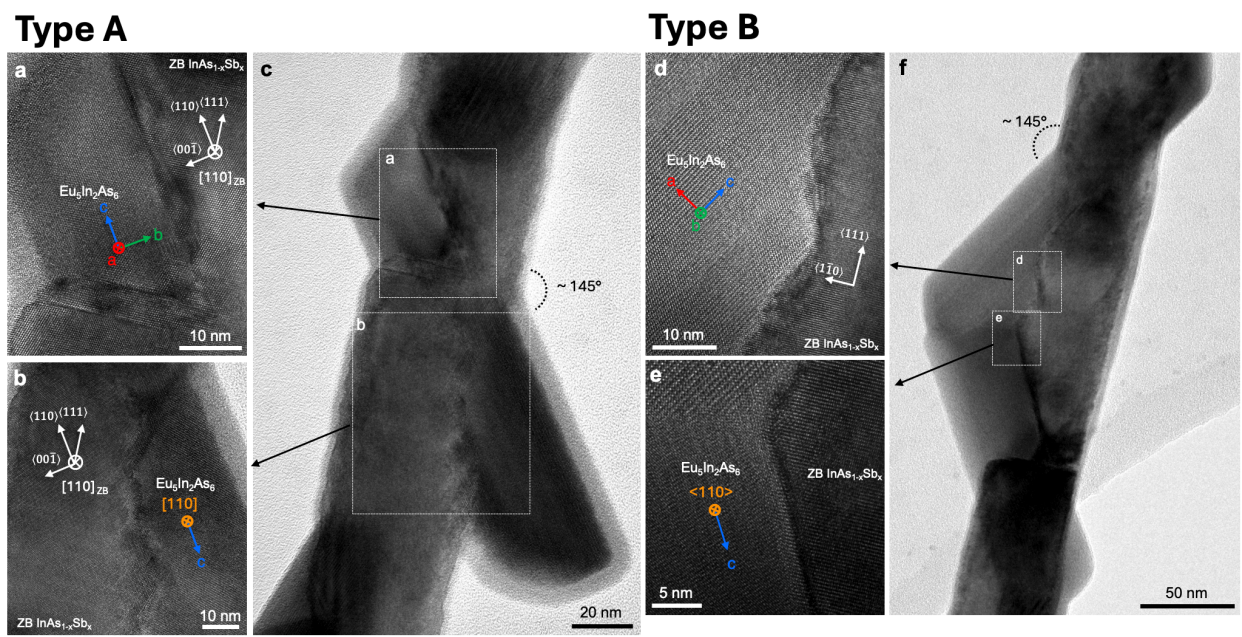

**Figure S02 Eu$_5$In$_2$As$_6$ grains along the [110].** (a–b) Magnified HR-TEM images of the interface of Type A between Eu$_5$In$_2$As$_6$ and InAsSb, taken from the dashed white rectangular areas in (c). (c) Low-magnification TEM image of Eu$_5$In$_2$As$_6$ grains on a ZB NW, corresponding to Figure 2d in the main text.

(d–e) Magnified HR-TEM images of the interface of Type B between $Eu_5In_2As_6$ and InAsSb, taken from the dashed white rectangular areas in (f). (f) Low-magnification TEM image of $Eu_5In_2As_6$ grains on a ZB NW, corresponding to Figure 2f in the main text.

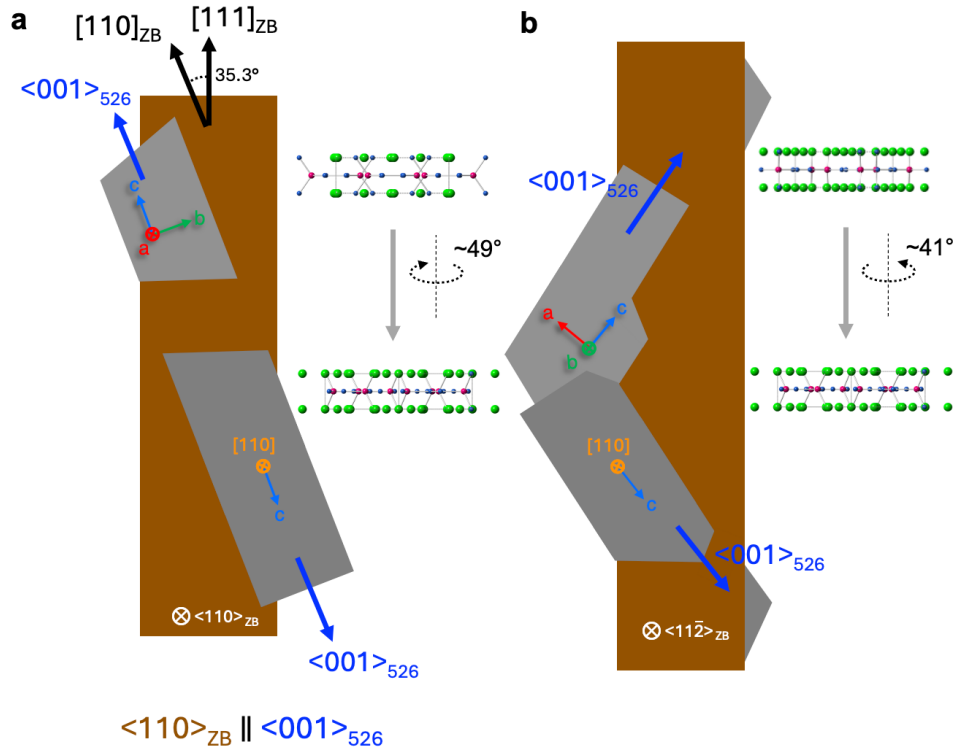

**Figure S03** Schematics illustrating the crystallographic directions for each material along the identical zone axes of the ZB structure: (a) <110>$_{ZB}$ and (b) <11-2>$_{ZB}$, based on Figures 2d and 2f in the main text, respectively. These schematics highlight the orientation relationships between the $Eu_5In_2As_6$ grains and the InAsSb NW cores, and the coexistence of another $Eu_5In_2As_6$ grain along the [110] in different zones.

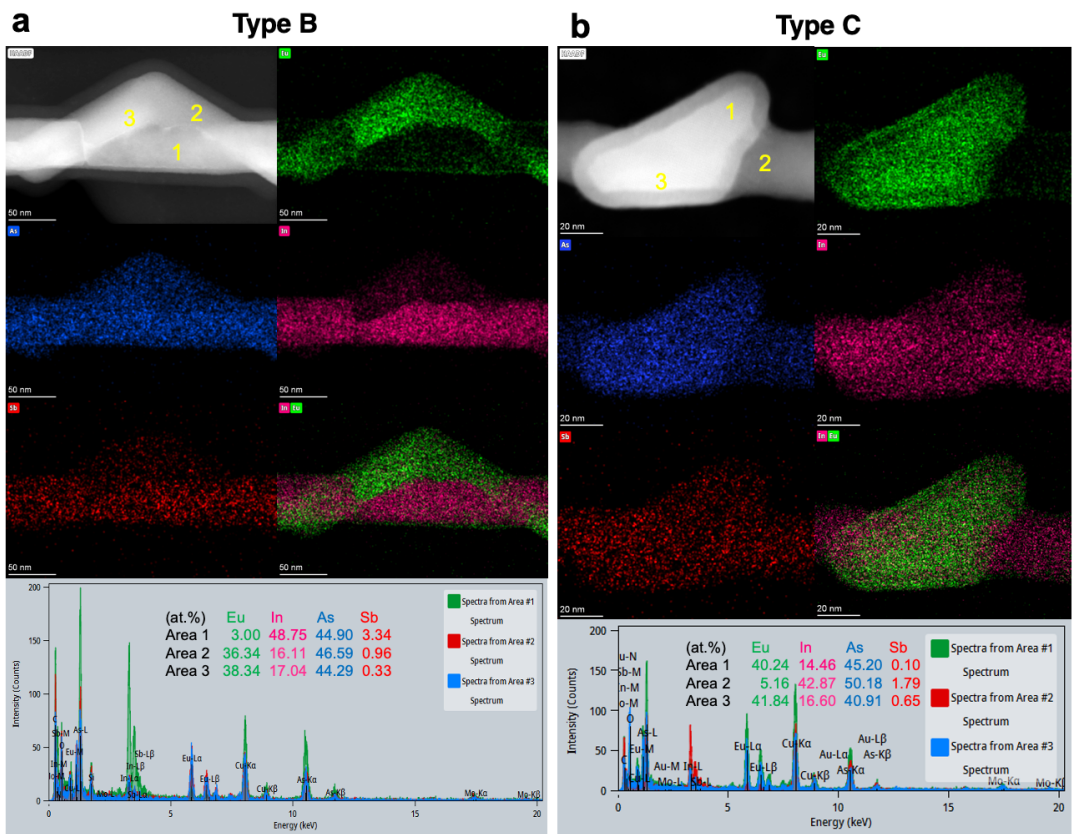

**Figure S04 Elemental composition of a Eu$_5$In$_2$As$_6$ grain and ZB InAsSb core in (a) Type B and (b) Type C, corresponding to Figures 2f and 2h in the main text, respectively. Top:** The Eu$_5$In$_2$As$_6$ and InAsSb areas were selected in HAADF STEM images for EDS measurement. EDS elemental maps show the distribution of Eu, In, As, Sb, and a composite map combining Eu and In. **Bottom:** Cumulative EDS spectra and their quantification extracted from the designated areas–Area 1, Area 2, and Area 3–in both types.

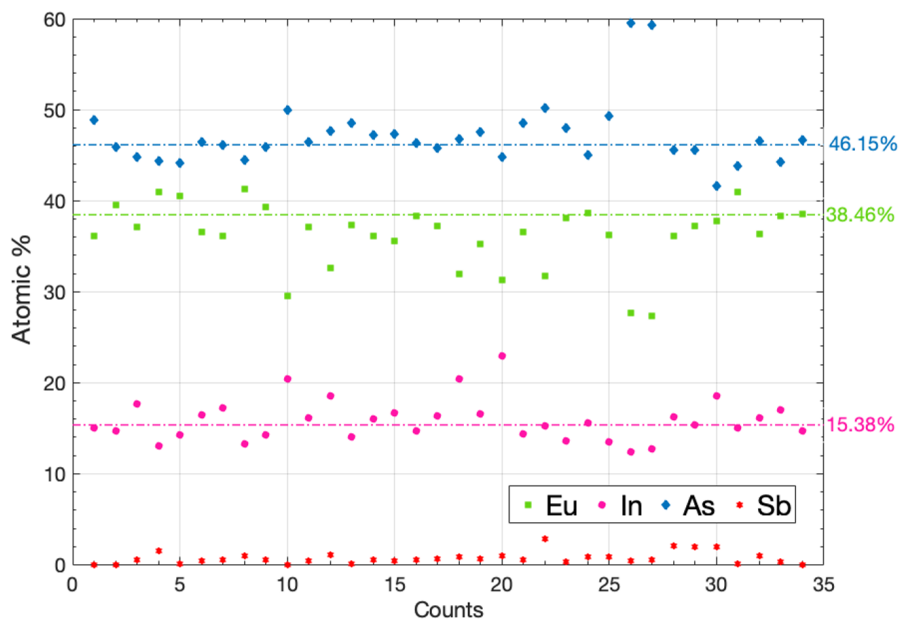

**Figure S05** EDS atomic percentage distribution of Eu (green square), In (pink circle), As (blue diamond), and Sb (red star) for selected $Eu_5In_2As_6$ grains was measured by STEM-EDS. The three dash-dotted guidelines represent the ideal stoichiometric composition of $Eu_5In_2As_6$.

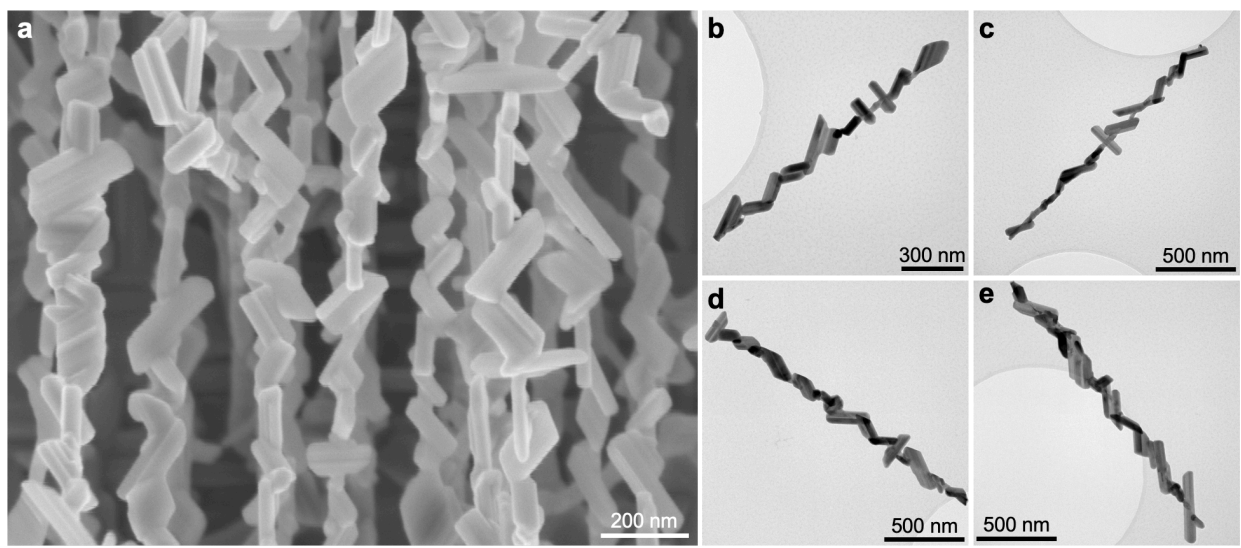

**Figure S06 Intense mutual exchange in reclined NWs on (001) substrates.** (a) Top-view SEM image and (b–e) TEM images of reclined $Eu_5In_2As_6$ NWs.

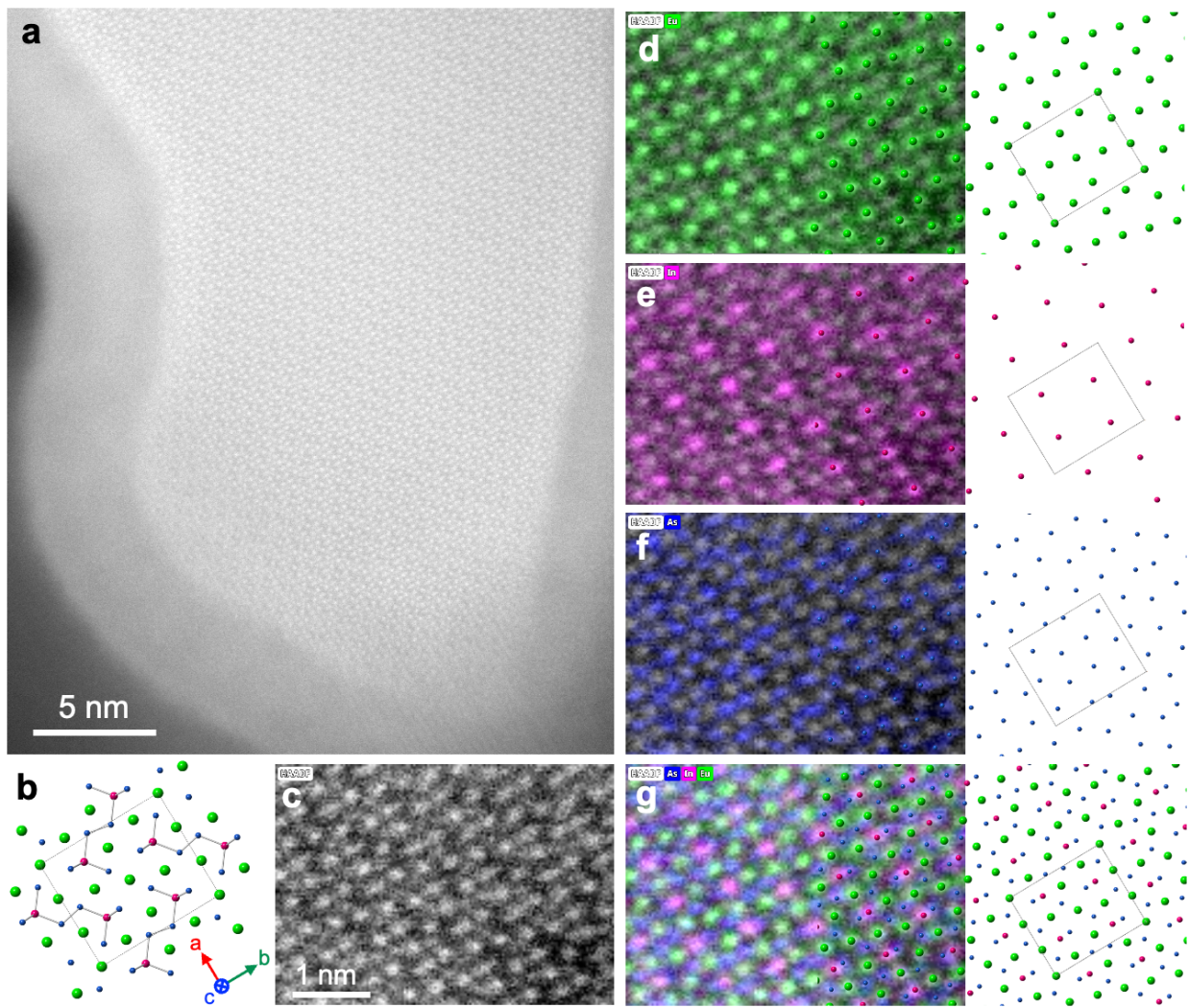

**Figure S07** (a) STEM-HAADF image of a type C $Eu_5In_2As_6$ grain. (b) Crystal structure of Zintl $Eu_5In_2As_6$. (c) Atomic-resolution STEM-HAADF image clearly shows the crystalline structure of $Eu_5In_2As_6$ along the <001> zone axis. (d–g) Corresponding EDS elemental maps of Eu, In, As, and the combined map of all three elements from (c), respectively. An atomic model is continuously overlaid on the right side. The dash-lined rectangle indicates a 1×1 unit cell of $Eu_5In_2As_6$ along the <001> zone axis, as depicted in (b).

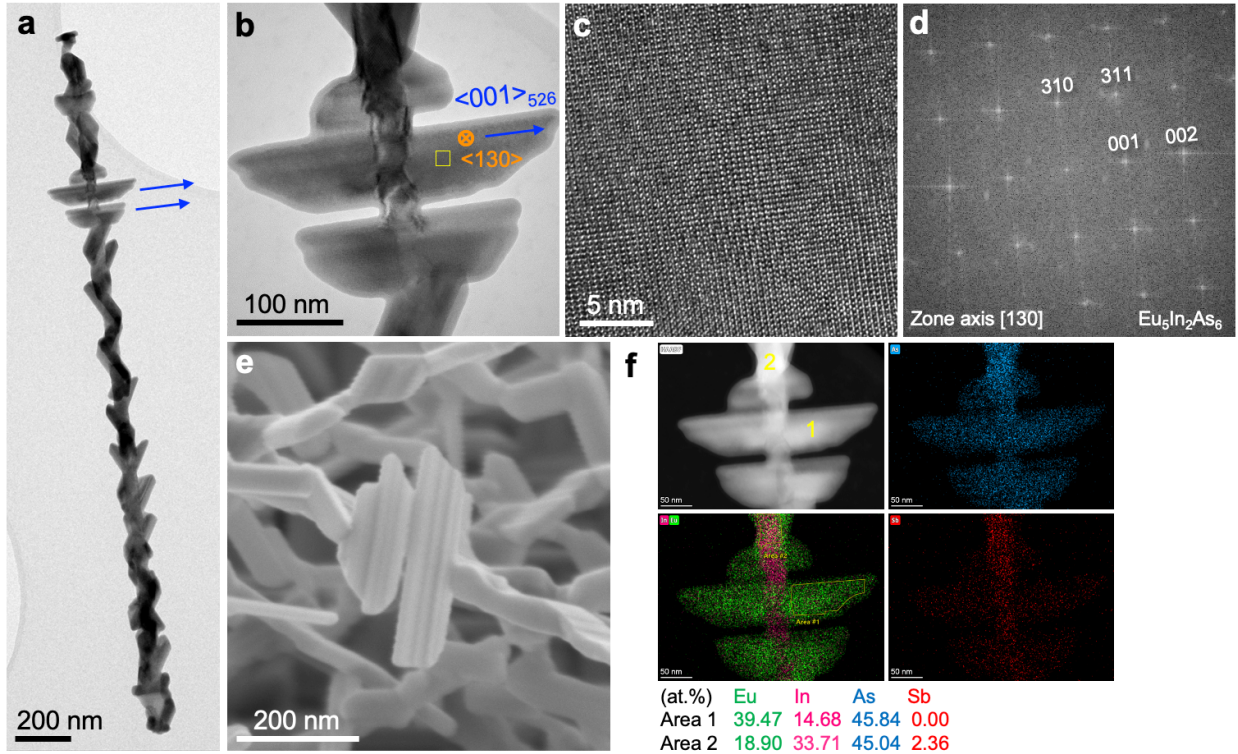

**Figure S08 Wing-shaped type C Eu$_5$In$_2$As$_6$ grains.** (a) TEM image of an Eu$_5$In$_2$As$_6$ nanowire. (b) Enlarged TEM image of the wing-shaped type C Eu$_5$In$_2$As$_6$ grains extending along the c-axis. (c) Further enlarged HRTEM image of the area indicated by the yellow square in (b). (d) FFT pattern from (c), showing the [130] zone axis of Eu$_5$In$_2$As$_6$. (e) A 45°-tilted SEM image of the wing-shaped grains. (f) HAADF (top left) and EDS elemental maps of As (top right), Sb (bottom right), Eu and In (bottom left) from (b). EDS spectra quantification for Areas 1 and 2 is shown below.

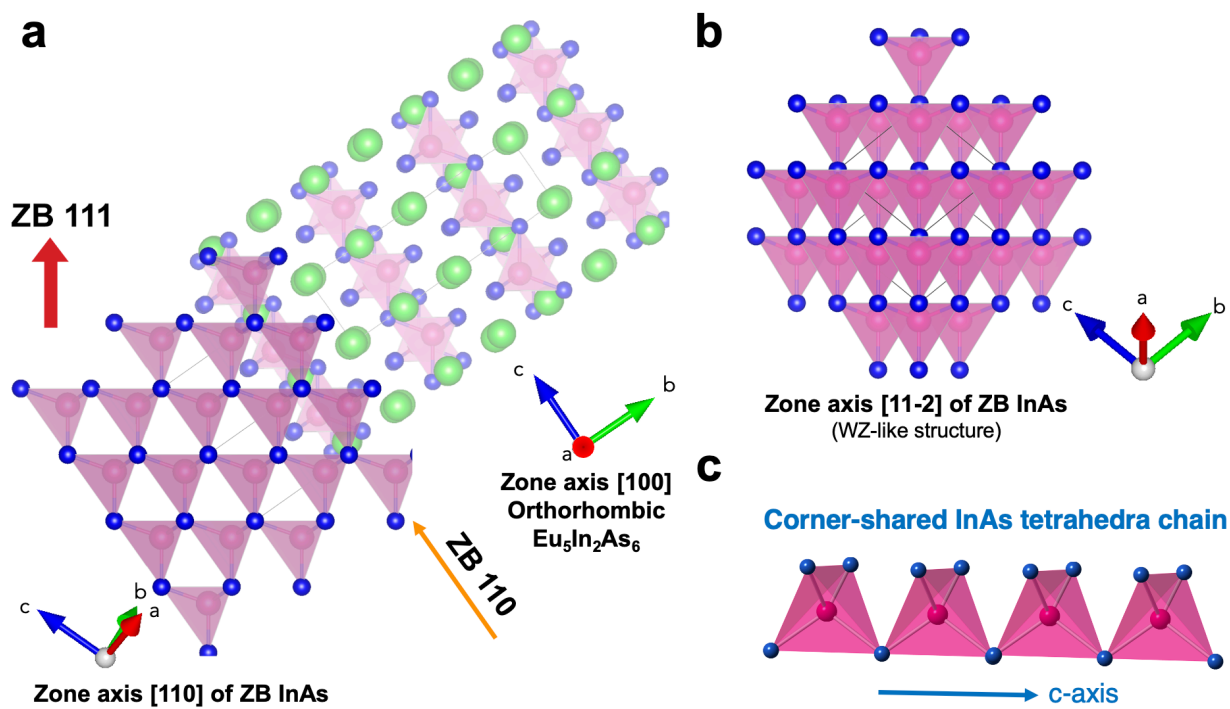

**Figure S09** (a) Atomistic model for the type A (and C) interface along the respective viewing directions. The left model represents a ZB InAs NW, growing along the <111> direction (upward). The orange arrow indicates the <110> direction of the boundary line, parallel to the c-axis of orthorhombic $Eu_5In_2As_6$. In $Eu_5In_2As_6$, the $[InAs_3]^{6-}$ chain consists of continuous corner-shared InAs tetrahedra aligned along the c-axis. (b) Atomistic model of InAs viewed along the [11-2] direction (type B), which is at a 30° relative to the <110> direction. From this perspective, the structure seemingly looks WZ InAs. (c) Schematic illustration of the corner-shared $[InAs_3]^{6-}$ chain in the Zintl phase $Eu_5In_2As_6$.

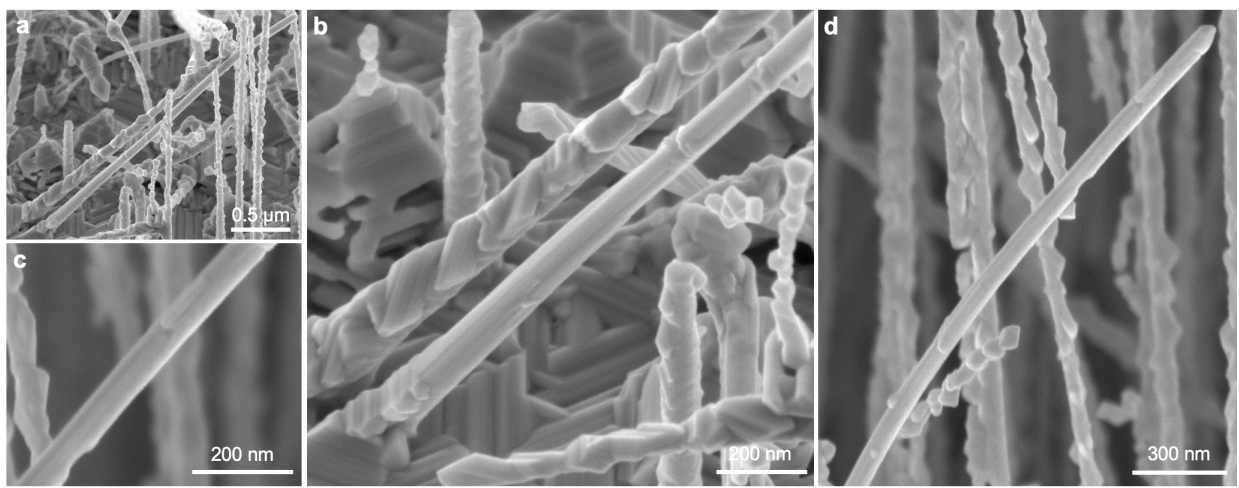

**Figure S10** (a–d) SEM images of $Eu_5In_2As_6$ smoothly coated NWs. Notably, the two NWs seen in (a) and (b) have clearly different surface morphologies.

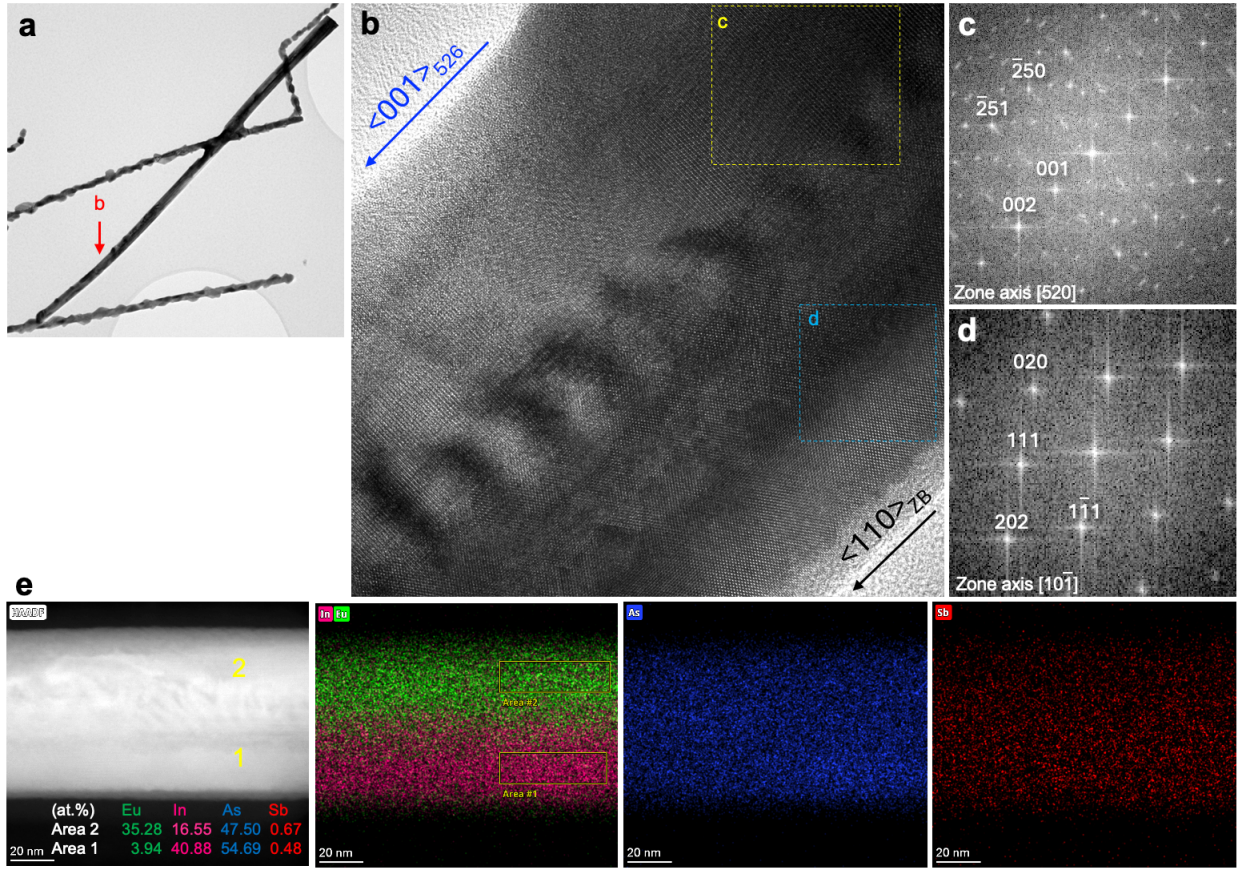

**Figure S11** (a) TEM image of smoothly coated $Eu_5In_2As_6$ NWs, identical to Figure 3b in the main text. (b) Enlarged TEM image near the tip indicated by the red arrow in (a), showing that the <001> direction of $Eu_5In_2As_6$ is parallel to the <110> direction of ZB InAsSb. (c) FFT pattern from the area indicated by the yellow rectangle in (b), revealing the [520] zone axis of $Eu_5In_2As_6$. (d) FFT pattern from the area indicated by the blue rectangle in (b), showing the [10-1] zone axis of ZB InAsSb (e) HAADF image and corresponding EDS elemental maps of Eu, In, As, and Sb from (b). The inset in HAADF image provides EDS spectra quantification for Area 1 (InAsSb) and Area 2 ($Eu_5In_2As_6$).

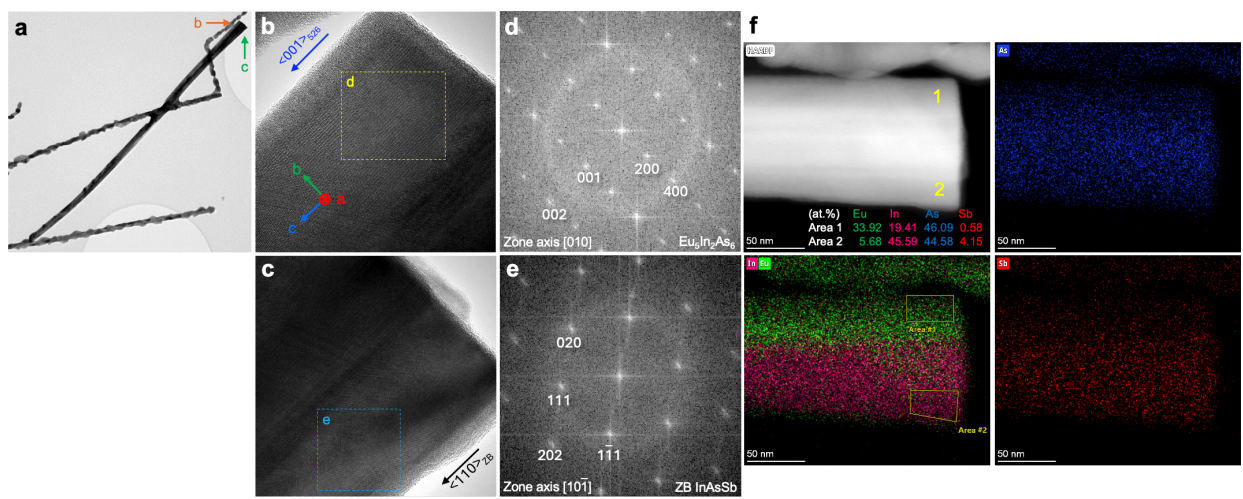

**Figure S12** (a) TEM image of Eu$_5$In$_2$As$_6$ NWs with a smooth coating, identical to Figure S10a. (b, c) Enlarged TEM images near the bottom region indicated by the red and green arrows in (a), respectively, showing that the <001> direction of Eu$_5$In$_2$As$_6$ is parallel to the <110> direction of ZB InAsSb. (d, e) FFT patterns from the areas indicated by the yellow and blue rectangles in (b) and (c) reveal the [010] zone axis of Eu$_5$In$_2$As$_6$ and the [10-1] zone axis of ZB InAsSb, respectively. (f) HAADF image and corresponding EDS elemental maps of As, In, Eu, and Sb from (b, c). The inset in the HAADF image provides EDS spectra quantification for Area 1 (InAsSb) and Area 2 (Eu$_5$In$_2$As$_6$).

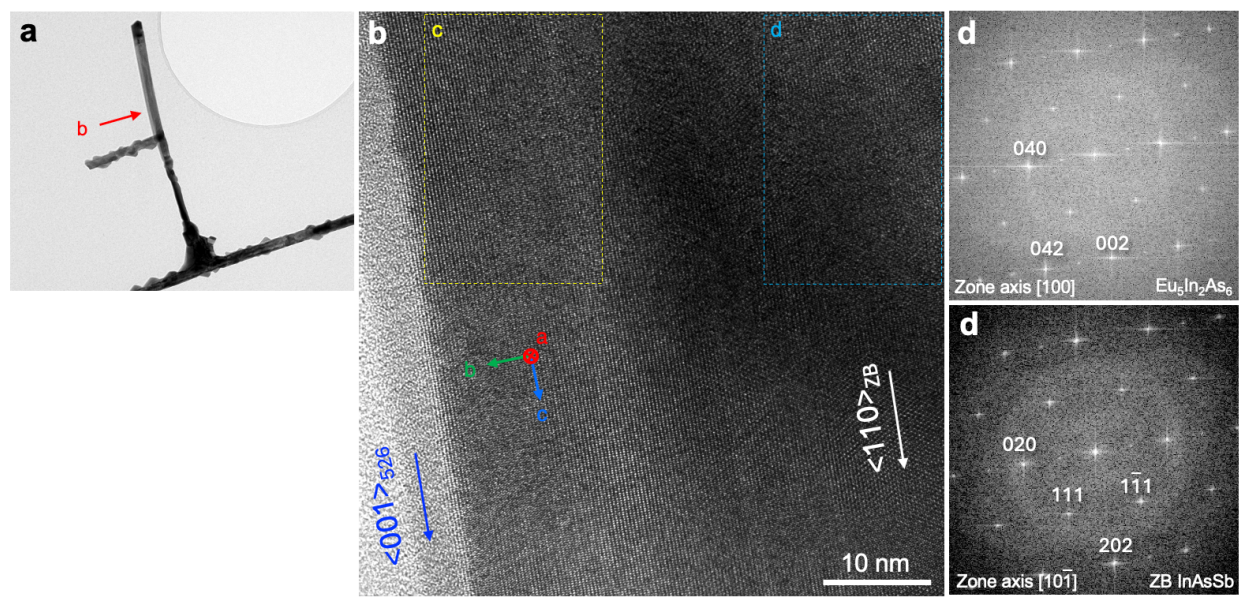

**Figure S13** (a) TEM image of smoothly coated Eu$_5$In$_2$As$_6$ NWs. (b) Enlarged HRTEM image from the area indicated by the red arrow in (a), showing that the <001> direction of Eu$_5$In$_2$As$_6$ is parallel to the <110> direction of ZB InAsSb. (d, e) FFT patterns from the areas indicated by the yellow and blue rectangles in (b) reveal the [100] zone axis of Eu$_5$In$_2$As$_6$ and the [10-1] zone axis of ZB InAsSb, respectively.

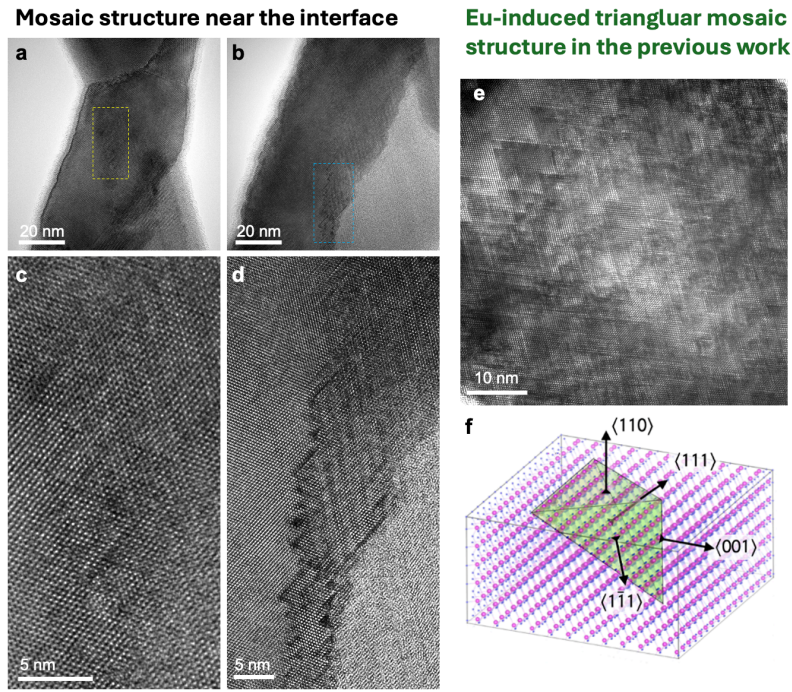

**Figure S14.** (a, b) TEM images of interface between the core NW and the Zintl grain along the <110> direction of ZB structure. (c, d) Enlarged TEM images of the areas indicated by the yellow and blue rectangles in (a) and (b), respectively, reveal a mosaic structure near the interface. (e) TEM image of the inversion domain boundary (IDB) network in the (EuIn)As NW[4]. (f) The 3D structure of the IDB is prismatic, bound by {111} planes[4]. The so-called mosaic structure in the ZB matrix results from the projection of a three-dimensional Eu atomic distribution.

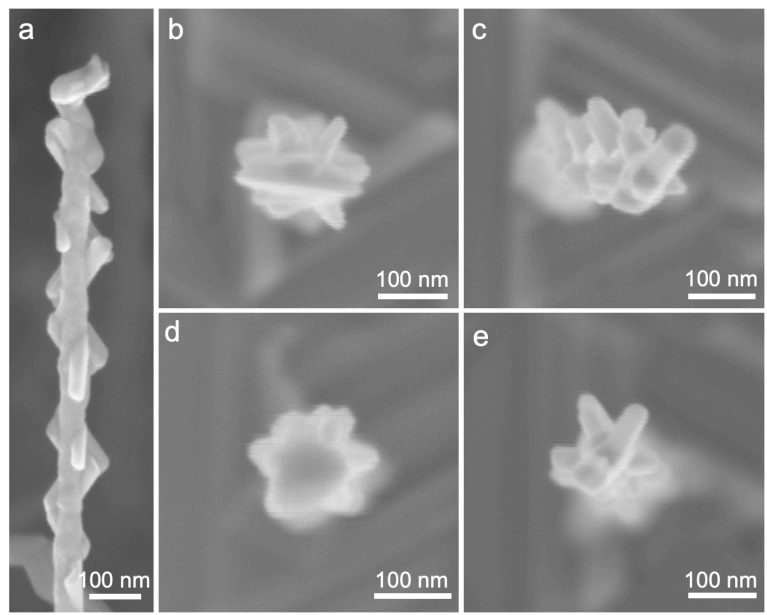

**Figure S15 Three-fold symmetry of $Eu_5In_2As_6$ grains.** (a) Side-view SEM image of an $Eu_5In_2As_6$ NW. (b–e) Top-view SEM images of the vertical $Eu_5In_2As_6$ NW.

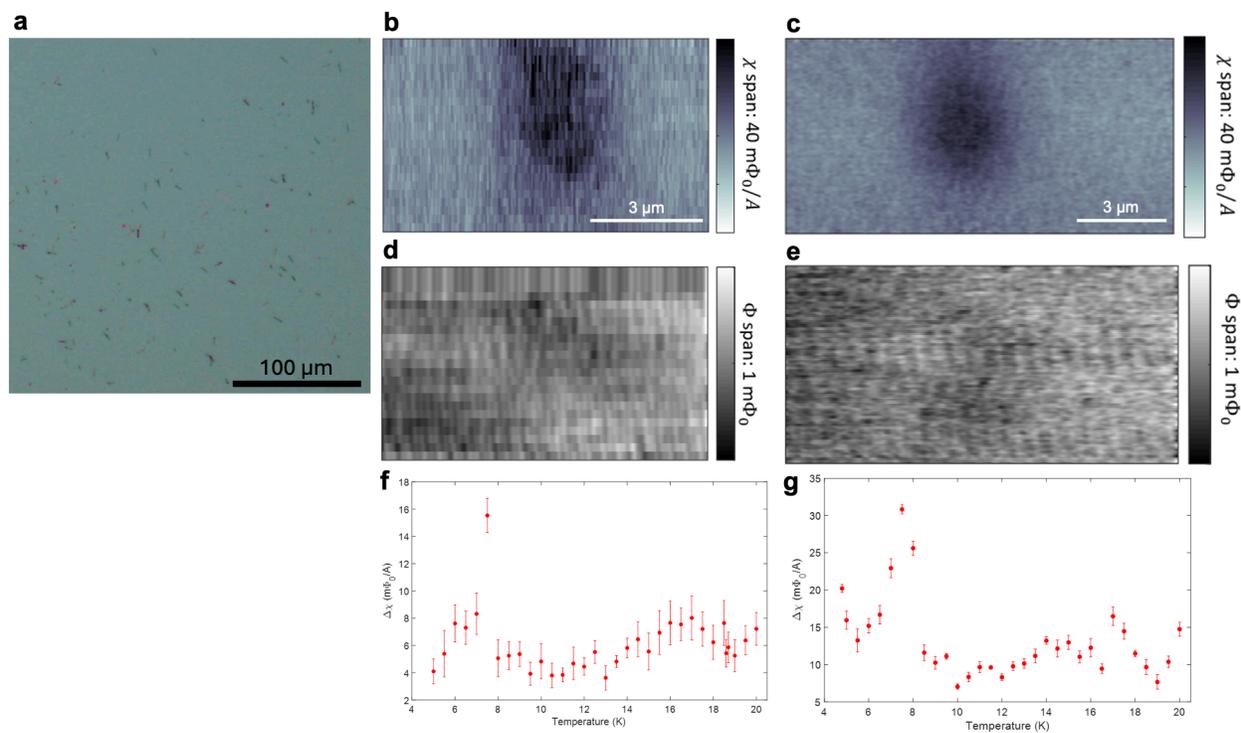

**Figure S16** (a) Optical microscopy image of $Eu_5In_2As_5$ NWs on a Si/SiO$_2$ substrate. (b, c) AC susceptibility maps of the region in (a), taken at 5 K. (d, e) DC magnetism maps corresponding to the same regions as (b) and (c), respectively. (f, g) Plots of AC susceptibility versus temperature for the NWs measured in (b) and (c), respectively. The error bars represent the standard deviation of pixel values in the relevant areas in the vicinity of the maximum local signal.

**Reference:**

(1) Kang, J.-H.; Cohen, Y.; Ronen, Y.; Heiblum, M.; Buczko, R.; Kacman, P.; Popovitz-Biro, R.; Shtrikman, H. Crystal Structure and Transport in Merged InAs Nanowires MBE Grown on (001) InAs. *Nano Lett.* **2013**, *13* (11), 5190–5196. https://doi.org/10.1021/nl402571s.

(2) Kang, J.-H.; Krizek, F.; Zaluska-Kotur, M.; Krogstrup, P.; Kacman, P.; Beidenkopf, H.; Shtrikman, H. Au-Assisted Substrate-Faceting for Inclined Nanowire Growth. *Nano Lett.* **2018**, *18* (7), 4115–4122. https://doi.org/10.1021/acs.nanolett.8b00853.

(3) Kang, J.-H.; Galicka, M.; Kacman, P.; Shtrikman, H. Wurtzite/Zinc-Blende 'K'-Shape InAs Nanowires with Embedded Two-Dimensional Wurtzite Plates. *Nano Lett.* **2017**, *17* (1), 531–537. https://doi.org/10.1021/acs.nanolett.6b04598.

(4) Shtrikman, H.; Song, M. S.; Załuska-Kotur, M. A.; Buczko, R.; Wang, X.; Kalisky, B.; Kacman, P.; Houben, L.; Beidenkopf, H. Intrinsic Magnetic (EuIn)As Nanowire Shells with a Unique Crystal Structure. *Nano Lett.* **2022**, *22* (22), 8925–8931. https://doi.org/10.1021/acs.nanolett.2c03012.

(5) Clarke, J.; I. Braginski, A. *The SQUID Handbook: Fundamentals and Technology of SQUIDs and SQUID Systems, I*; John Wiley & Sons, Ltd, 2004.

(6) Persky, E.; Sochnikov, I.; Kalisky, B. Studying Quantum Materials with Scanning SQUID Microscopy. *Annual Review of Condensed Matter Physics* **2022**, *13* (1), 385–405. https://doi.org/10.1146/annurev-conmatphys-031620-104226.

(7) Song, M. S.; Houben, L.; Zhao, Y.; Bae, H.; Rothem, N.; Gupta, A.; Yan, B.; Kalisky, B.; Zaluska-Kotur, M.; Kacman, P.; Shtrikman, H.; Beidenkopf, H. Topotaxial Mutual-Exchange Growth of Magnetic Zintl Eu3In2As4 Nanowires with Axion Insulator Classification. *Nat. Nanotechnol.* **2024**, *19* (12), 1796–1803. https://doi.org/10.1038/s41565-024-01762-7.